\pdfoutput=1

\documentclass[11pt,twoside,a4paper,cmspaper,final,collab]{cms-tdr}
\def\svnVersion{258629}\def\cmsCernNo{2013-037}\def\cmsCernDate{\today}\def\cmsMessage{Published in Physical Review C as \href{http://dx.doi.org/10.1103/PhysRevC.90.024908}{\doi{10.1103/PhysRevC.90.024908.}}}

\begin{document}\cmsNoteHeader{HIN-12-013}

\hyphenation{had-ron-i-za-tion}
\hyphenation{cal-or-i-me-ter}
\hyphenation{de-vices}
\RCS$Revision: 251535 $
\RCS$HeadURL: svn+ssh://svn.cern.ch/reps/tdr2/papers/HIN-12-013/trunk/HIN-12-013.tex $
\RCS$Id: HIN-12-013.tex 251535 2014-07-16 12:01:39Z alverson $

\newlength\cmsFigWidth\setlength\cmsFigWidth{0.95\textwidth}

\providecommand{\pp}{\Pp\Pp\xspace}
\providecommand{\PbPb}{\mbox{PbPb}\xspace}
\providecommand{\HYDJET}{\textsc{hydjet}\xspace}
\providecommand{\PYHYD}{\textsc{pythia}+\textsc{hydjet}\xspace}
\providecommand{\sNN}{\ensuremath{\sqrt{\smash[b]{s_{_\mathrm{NN}}}}}\xspace}
\ifthenelse{\boolean{cms@external}}{\providecommand{\cmsTableResize[1]}{\relax{#1}}}{\providecommand{\cmsTableResize[1]}{\resizebox{\columnwidth}{!}{#1}}}
\cmsNoteHeader{HIN-12-013}
\title{Measurement of jet fragmentation in \texorpdfstring{\PbPb and \pp collisions at $\sNN=2.76$\TeV}{PbPb and pp collisions at sqrt(s[NN]) = 2.76 TeV}}

\date{\today}

\hyphenation{mis-i-den-ti-fied}%

\abstract{
The jet fragmentation function of inclusive jets with transverse momentum \pt above 100\GeVc in \PbPb collisions has been measured using reconstructed charged particles with \pt above 1\GeVc in a cone of radius 0.3 around the jet axis. A data sample of \PbPb collisions collected in 2011 at a nucleon-nucleon center-of-mass energy of $\sNN = 2.76\TeV$ corresponding to an integrated luminosity of 150\mubinv is used. The results for \PbPb collisions as a function of collision centrality and jet transverse momentum are compared to reference distributions based on \pp data collected at the same center-of-mass energy in 2013, with an integrated luminosity of 5.3\pbinv. A centrality-dependent modification of the fragmentation function is found. For the most central collisions, a significant enhancement is observed in the PbPb/\pp fragmentation function ratio for charged particles with \pt less than 3\GeVc. This enhancement is observed for all jet \pt bins studied.
}

\hypersetup{%
pdfauthor={CMS Collaboration},%
pdftitle={Measurement of jet fragmentation in PbPb and pp collisions at sqrt(s[NN]) = 2.76 TeV},%
pdfsubject={CMS},%
pdfkeywords={CMS, Heavy Ion Physics, Jet fragmentation function, Jet Tracks}}
\maketitle
\section{Introduction}

High-energy collisions of heavy ions provide an important experimental tool
to study the puzzles of confinement and chiral symmetry breaking in
Quantum Chromodynamics (QCD), the theory of strong interactions.
It is expected that a state of deconfined and chirally symmetric quarks and gluons, called the quark-gluon plasma (QGP), exists in the hot and dense QCD medium produced in heavy-ion collisions \cite{Shuryak:1977ut,Collins:1974ky,Cabibbo:1975ig,Freedman:1976ub}.
However, due to the complexity of multi-body collision systems, finding
clean experimental signatures of the QGP poses a challenge.
In 1982,  Bj\"orken first conjectured that
high-energy partons produced in nucleus-nucleus collisions
would lose energy as they traverse the QGP~\cite{Bjorken:1982tu}.
Since then, a variety of experimental observations,
including the suppression of high transverse momentum (\pt) particles, have provided evidence for this conjecture. This suppression was first seen at RHIC~\cite{Adcox:2001jp,Adler:2002xw} and later at the CERN LHC (see, for example, Ref.~\cite{CMS:2012aa} and references therein).
In addition, the analysis of the first data from the CERN LHC provided more direct evidence of parton energy loss
using the difference in \pt
 between back-to-back pairs of jets \cite{Collaboration:2010bu,CMS_dijet2010,CMS_dijet} and also jet-photon pairs~\cite{Chatrchyan:2012gt}.
Unbalanced dijet and jet-photon pairs were found to be much more prevalent in
the most central \PbPb collisions~\cite{CMS_dijet2010,CMS_dijet, Chatrchyan:2012gt}
compared to expectations in the absence of a hot and dense medium.
Further theoretical QCD studies have been inspired by these observations using jets from the CERN LHC.
On the other hand, it has been predicted that in the presence of the strongly interacting medium produced in heavy-ion collisions, the partitioning of the parton energy into particles (the fragmentation function) may be modified and the yield of high-\pt{} particles suppressed~\cite{Wang:2001ifa,Renk:2008pp,Renk:2009nz,Armesto:2009fj,Wang:2009qb}.
Therefore, direct measurements of jet fragmentation
in heavy-ion collisions are important in the quest for understanding QCD
through medium-induced parton energy loss.

Studies of the detailed jet structure also have
important practical consequences for other aspects of jet analyses, including
the connection to the kinematics of the partons that produce jets.
The longitudinal and transverse fragmentation properties of jets
connect the perturbatively calculable production of high-\pt quarks
and gluons with the hadronized final-state particles. The study of
jet production requires reconstructing the jets using final-state particles.
This reconstruction relies on hadronization
models to quantify how the original jet energy is related to the
energy determined by adding the energies of the individual particles.
In addition, although the production cross section can be calculated perturbatively, there are corrections due to the non-perturbative hadronization
process.
Study of the fragmentation function
provides an important experimental check on the validity of the
assumed jet fragmentation in heavy-ion collisions.
These results can also be used
to directly connect jet observables to measurements of high-\pt
particles.

The goal of this analysis is to measure the jet fragmentation function in heavy-ion collisions using reconstructed jets. This is in contrast to the parton fragmentation function measured in $\Pep\Pem$ experiments, which is obtained relative to the initial parton momentum.
In previous work~ \cite{Chatrchyan:2012gw},
the higher \pt ($\pt>4\GeVc$) component of the fragmentation function
was found to be qualitatively similar to that for jets in \pp collisions, for which the medium is absent.
The analysis described in this paper uses data from the 2011 \PbPb\ run at a nucleon-nucleon center-of-mass energy of $\sNN=2.76$\TeV.
This work expands on the previous results
by measuring the fragmentation function for particles down to \pt of 1\GeVc. Taking advantage of the higher integrated luminosity (150 \mubinv for \PbPb and 5.3\pbinv for the \pp reference data), the measurement is also carried out in more differential centrality bins, and as a function of jet \pt.
This measurement complements the previously published observation of a modification of the transverse profile of the jet in \PbPb \cite{Chatrchyan:2013kwa}, using the same 2011 and 2013 data.

\section{The CMS detector}

The centerpiece of the CMS detector is a superconducting solenoid,
12.5\unit{m} long with an internal diameter of 6\unit{m}, that provides a uniform magnetic field of
3.8\unit{T}.
In the CMS coordinate system,  the $z$ axis
points in the counterclockwise beam direction,
the $x$ axis points towards the centre of the LHC ring,
and the $y$ axis points up, perpendicular to the plane of the LHC ring.
The azimuthal angle $\phi$ is measured
with respect to the $x$ axis, and the polar angle $\theta$ is measured with respect to the $z$ axis.
Charged particles or charged particles reconstructed
in the
inner tracking system are characterized by their
transverse momentum, $\pt = \abs{\vec{p}}\,\sin{\theta}$,
and pseudorapidity, $\eta = -\ln\left[ \tan(\theta/2) \right]$.
The inner tracking system is composed of a pixel detector with
three barrel layers at radii between 4.4 and 10.2\unit{cm}
and a silicon strip tracker with 10 barrel layers extending
outwards to a radius of 110\unit{cm}.  Two
endcap modules extend the acceptance of the tracking system up to $\abs{\eta}=2.5$.
The momentum resolution for reconstructed tracks
in the barrel region is about 1\% at \pt = 100\GeVc and up to 2\% in the endcap at the same \pt.

The calorimeters inside the magnetic coil consist of a
lead-tungstate crystal electromagnetic calorimeter (ECAL)  and a
brass/scintillator hadron calorimeter (HCAL) with coverage up
to $\abs{\eta}=3$.
Steel/quartz-fibre
Cherenkov hadron forward (HF) calorimeters extend the coverage to $\abs{\eta}$ = 5.2.
Muons are measured in gas-ionization detectors embedded in the steel
flux-return yoke of the magnet.
The calorimeter cells are
grouped in projective towers of granularity
$\Delta\eta\times\Delta\phi = 0.087 \times 0.087 $
for the central rapidities ($\abs{\eta}\le 2$) considered in this paper.
The energy scale in data agrees with that in the simulation to better than 1\% in the barrel region ($\abs{\eta}<1.5$)
and better than 3\% in the endcap region ($1.3<\abs{\eta}<3.0$)~\cite{Chatrchyan:2013dga}.
Hadron calorimeter cells in the $\abs{\eta}<3$ region are calibrated primarily with test-beam data and radioactive sources~\cite{hcal_jinst,HcalTestBeam}.
A detailed description of the CMS detector can be found in Ref.~\cite{CMS_Detector}.

\section{Trigger, event selection, and centrality determination}

\label{sec:evtsel}

The CMS high-level trigger (HLT) system is used to select \PbPb collision events containing high-\pt jets reconstructed from calorimeter towers. The trigger threshold applied to the jet transverse momentum is \pt = 80\GeVc. For \pp collisions, the threshold
is 60\GeVc.
In addition to the online trigger decision, standard offline selection criteria are applied to remove backgrounds due to detector noise, beam gas collisions, beam scraping and
ultra-peripheral-collision events~\cite{CMS_dijet}. Events are further restricted to those with a reconstructed vertex which
includes at least two tracks and has a $z$ position within 15\unit{cm}of the detector center. Finally, an offline HF coincidence of
at least three towers with energy greater than 3\GeV on each side of the interaction point is required. These event-quality requirements have only a small effect on the number of selected events, and have a negligible impact on the jet analyses~\cite{CMS_dijet,Chatrchyan:2012gt}.

For the analysis of \PbPb data, it is important to determine the collision centrality in each event, which is related to the overlap between the
two colliding nuclei. Centrality is determined using
the sum of transverse energy (energy times $\sin(\theta)$) reconstructed in the HF (covering $2.9<\abs{\eta}<5.2$). The HF transverse energy distribution is used
to divide the event sample into percentiles of the total nucleus-nucleus hadronic interaction cross section.
A detailed description of the centrality determination can be found in~\cite{CMS_dijet}.

\section{Monte Carlo simulations}
\label{sec:mc}
In this analysis, Monte Carlo (MC) simulations have been used primarily for evaluation of reconstruction performance, particularly in determination of tracking efficiency, and jet energy response and resolution. These studies mostly used QCD jet events simulated by the \PYTHIA MC generator~\cite{Pythia} (version 6.423, tune Z2)~\cite{Field}. These simulated \PYTHIA events are propagated through the CMS detector using the \GEANTfour package~\cite{GEANT4} to simulate the detector response.

In order to account for the effect of the \PbPb\ underlying event, the \PYTHIA events are embedded into fully simulated \PbPb\ events, generated by the \HYDJET event generator~\cite{Lokhtin:2005px} (version 1.8) which has been tuned to describe the centrality dependence of the hadron multiplicity, \pt spectra, and elliptic flow in minimum bias \PbPb\ data. The embedding
is done by mixing the simulated digital information from \PYTHIA and \HYDJET.

\section{Reconstruction}
\subsection{Jet reconstruction}
\label{sec:jetrecon}

For both \pp and \PbPb collisions, the analysis is based on jets reconstructed using the anti-\kt algorithm, with a distance parameter $R$ of 0.3, utilizing particle-flow (PF) objects that combine tracking and calorimetric information~\cite{pas_ParticleFlow,CMS-PAS-PFT-10-002}. The value of 0.3 for $R$, also used in previous CMS jet studies \cite{CMS_dijet, Chatrchyan:2012gt}, was chosen to optimize the jet reconstruction efficiency and misidentification rate. In the \PbPb data, the contribution of the underlying heavy-ion event is removed using an iterative pileup subtraction method~\cite{Kodolova_JetReco}.

Only events with a minimum jet $\pt > 100\GeVc$, for which the trigger is more than 99\% efficient, are selected.  From this event sample, all of the PF jets above $\pt >100\GeVc$ and also within $0.3<\abs{\eta}<2$ are included in the analysis.
Jets in the central region of $\abs{\eta}<0.3$ are excluded in order to avoid overlaps in the background subtraction procedure (see Section~\ref{sec:UE}).

\subsection{Charged-particle reconstruction}

Charged particles are reconstructed in the following steps, which are similar to those used previously~\cite{CMS:2012aa,Chatrchyan:2013kwa}. First, using a collection of pixel-only tracks required to have three hits in the pixel detector, the three-dimensional primary vertex position is estimated from extrapolations back to the beam spot region.
Next, track candidates are built from triplet seeds, consisting of hits in three layers of the pixel barrel and endcap detectors.
To reduce the random combinatorial background, only seeds pointing to a restricted region within 2\unit{mm} of the primary vertex and with a minimum \pt of 0.9\GeVc are used. Further selections are made on
the normalized goodness of the track fit (\ie $\chi^2$ per degree of freedom) and on the compatibility of the fitted
triplet seeds with the primary vertex, before propagating the seed trajectories through the strip tracker
to build fully reconstructed tracks. To improve the track reconstruction efficiency, two more iterations of
the tracking are performed after removing hits unambiguously belonging to the tracks found in the first
iteration. This procedure is based on the standard \pp iterative tracking~\cite{Khachatryan:2010pw}. More efficient
pp-based triplet track and pixel pair seedings are used in the second and third iterations, respectively.
The tracks found in the later iterations are merged with the first-iteration tracks after removing any
duplicate tracks, based on the fraction of shared hits.
In all iterations, track quality criteria are applied to the final list of track candidates to reduce the reconstruction misidentification rate.
The minimal \pt for tracks used in the analysis is 1\GeVc.

A complete understanding of the tracking performance is of primary importance for this analysis. This was studied using jet events simulated with \PYTHIA (tune Z2) embedded into a {\HYDJET} 1.8 background. The track-by-track
corrections for reconstruction efficiency and misidentified tracks are computed in bins of track $\eta$ and $\pt$, neighboring jet $\pt$, and event centrality without
any selection criteria imposed on the reconstructed jets.

The performance of the tracking algorithm depends on the local environment in which it operates. Therefore, the
corrections are computed separately for the four centrality classes used in the analysis: 0--10\%, 10--30\%, 30--50\%,  and 50--100\%, (most central to most peripheral events).  Due to the low multiplicity in the 50--100\% centrality, the correction is also used for the split 50--70\% and 70--100\% centrality bins, for the five centrality class results.
At low
$\pt$, the efficiency is $\approx$10\% higher for the pure {\PYTHIA} sample (\ie \pp multiplicity environment) than for the most central {\HYDJET}-embedded \PYTHIA (\ie \PbPb multiplicity environment), while at high $\pt$ the difference is about 4\%. The misidentification rate
is small for all samples and ranges from 4\% at 1\GeVc to 2\% at 120\GeVc.

\section{Analysis}

\subsection{Jet fragmentation function} \label{sec:jetFF}

The jet fragmentation function is measured by correlating reconstructed
charged-particle tracks contained within the jet cones, with the axis of the respective jet~\cite{PhysRevD.64.032001}.
As done in previous measurements at hadron colliders \cite{PhysRevLett.65.968,PhysRevD.68.012003},
the fragmentation function is presented as a function of the variables $z$ and $\xi$, defined as
\begin{equation*}
\label{eq:xi}
z = \frac{p_{\parallel}^{\text{track}}}{p^{\text{jet}}}, \quad \xi = \ln \frac{1}{z},
\end{equation*}
where $p_{\parallel}^{\text{track}}$ is the momentum component of the track
along the jet axis and $p^{\text{jet}}$ is the magnitude of the jet momentum.
All tracks in a cone of $\sqrt{\smash[b]{(\Delta\phi)^2+(\Delta \eta)^2}} < 0.3$ around the jet axis are included in the analysis.
The fragmentation function, defined as $\left( 1/N_\text{jet}\right )\ \rd{}N_{\text{track}}/\rd\xi$, is
normalized to the total number of jets ($N_\text{jet}$).

\subsection{Underlying event subtraction}
\label{sec:UE}

Due to the high level of underlying event activity coming from the heavy-ion collisions,
tracks that are not associated with the jet fragmentation can be found  within the jet cone.
This uncorrelated background contribution to the fragmentation function is subtracted statistically using the so-called $\eta$-reflection method.
The underlying event contribution
is estimated by selecting charged particles that lie in a ``background'' jet cone obtained by reflecting the original jet cone around $\eta = 0$ while keeping the same $\phi$ coordinate.
The background distribution is then subtracted from the raw distribution obtained from tracks in the jet cone.
The use of this procedure is the reason that jets in the region $\abs{\eta}< 0.3$ are excluded to avoid overlap between the signal jet region and the region used for background estimation.
\subsection{Fragmentation function biases}
\label{sec:biasCorrections}

The measured fragmentation function contains two sources of bias introduced by the jet reconstruction.
The first bias results from jet reconstruction in the
presence of a high-multiplicity background.
As a result of the steeply falling jet spectrum,
a significant fraction of reconstructed  jets that just barely exceed the minimum \pt threshold
come from the more abundant lower-energy jets enhanced by
an upward fluctuation in the background.
The second bias is caused by the jet reconstruction being correlated with the jet's fragmentation pattern. Jets which fragment harder, \ie those that produce fewer particles but with higher average \pt, are easier to reconstruct and have a slightly higher energy scale.

Both biases are corrected for in the final
analysis. The biases affect different parts of the fragmentation function and are assumed to be independent. The correction factors are individually derived based on the {\PYHYD} simulation.
The first bias is corrected by comparing the estimated background in the reflected cone
to the true background in the jet cone. A correction factor is then derived based on the ratio of the two and used to weight tracks in the background cone.
The correction is computed as a function of track \pt and applied to events in the two highest analysis centrality classes where background tracks dominate over the signal tracks at low \pt.
An important element of this MC-based technique is the fact that the tracks in the background cone have very similar fluctuations (\ie similar variations in background energy) in the simulated heavy-ion events and in the data.
The second bias is corrected by comparing the
true fragmentation function of reconstructed jets in \PYTHIA signal events and
\PYHYD events. A ratio is derived based on the two and used to weight tracks in the \pp jet cone so that the \pp reference data can be consistently compared with PbPb. This correction is cross-checked by repeating the full analysis for different kinds of signal events with embedded jets, as described in Section~\ref{sec:trackeff_syst}.
\subsection{Proton-proton reference data}
\label{sec:ppsmearing}
\label{sec:jetFFanalysisPP}

In order to quantify any medium-related effects, the results are compared to reference distributions using the high statistics jet data in \pp collisions collected in 2013 at $\sNN=2.76$\TeV corresponding to an integrated luminosity of $5.3\pm 0.2$\pbinv.
For a direct comparison between \pp and \PbPb collisions, the jet momentum
resolution deterioration in \PbPb\ events has to be taken into account.
For this purpose, the reconstructed \pt\ of every jet in the \pp data is
smeared using a Gaussian distribution based on the quadratic difference of the jet momentum resolution
in \PbPb\ and \pp\ data.
The jet momentum resolutions are derived from the \PbPb and \pp MC simulations described in Section~\ref{sec:mc}.
In order to keep the jet kinematic constraints consistent,
a reweighting factor, derived based on the ratio of the \PbPb{} and the smeared \pp jet \pt spectra, is applied to each \pp jet.
After the reweighting procedure, the resulting \pp jet \pt distribution matches the one in \PbPb in each centrality bin of the analysis.
\section{Systematic uncertainties} \label{sec:systematics}

\label{sec:trackeff_syst}

\begin{table*}[tb]
\centering
\topcaption{Summary of systematic uncertainties in jet fragmentation function
  analysis in bins of $\xi$ for the 0--10\% centrality.  The values indicate the typical values of the systematic uncertainties and the intervals indicate the range of systematic uncertainties for various ranges of $\xi$.}
  \label{table:JetFFSys}

 \cmsTableResize{ \begin{scotch}{c|c|cccc}

Item & Input/variation & \multicolumn{4}{c}{$\xi<1.5$}\\
\hline
\multicolumn{2}{c|}{Jet \pt range (\GeVcns)} & 100--120 & 120--150 & 150--300 & 100--300\\
\hline

Jet energy resolution & 10--20\% smearing & 3.7--6.7\% & 2--6.2\% & 3.9--6.9\% & 0.1--0.7\%\\
Jet energy scale & 5\% shift & 9.3--29\% & 8.5--26\% & 7.9--25\% & 8.9--28\%\\
Tracking efficiency & Non-closure & \multicolumn{4}{c}{0.1--1.9\%}\\
Tracking efficiency & Centrality variation & \multicolumn{4}{c}{10\%}\\
Background bias &50\% of correction & \multicolumn{4}{c}{$<$0.1\%}\\
Background procedure & Diff. between methods & \multicolumn{4}{c}{1.9--4.3\%}\\
Gluon jet cross-check & Non-closure & \multicolumn{4}{c}{1.2--3.9\%}\\
Quark jet cross-check & Non-closure & \multicolumn{4}{c}{4.8--9.8\%}\\
\hline
Total & & 15--28\% & 14--37\% & 14--20\% & 15--31\%\\

\hline\hline
Item & Input/variation & \multicolumn{4}{c}{$1.5<\xi<4$}\\
\hline
\multicolumn{2}{c|}{Jet \pt range (\GeVcns)} & 100--120 & 120--150 & 150--300 & 100--300\\
\hline

Jet energy resolution & 10--20\% smearing & 0.1--2.7\% & 1.1--1.4\% & 0--1.9\% & 0.1--1.0\%\\
Jet energy scale & 5\% shift & 1.6--4.8\% & 1.4--4.4\% & 1.3--4.1\% & 1.5--4.6\%\\
Tracking efficiency & Non-closure & \multicolumn{4}{c}{0.1--0.7\%}\\
Tracking efficiency & Centrality variation & \multicolumn{4}{c}{10\%}\\
Background bias &50\% of correction & \multicolumn{4}{c}{0--3.3\%}\\
Background procedure & Diff. between methods & \multicolumn{4}{c}{0--2.7\%}\\
Gluon jet cross-check & Non-closure & \multicolumn{4}{c}{3.8--5.0\%}\\
Quark jet cross-check & Non-closure & \multicolumn{4}{c}{0.2--2.0\%}\\
\hline
Total & & 11--14\% & 11--13\% & 11--13\% & 11--12\%\\

\hline\hline
Item & Input/variation & \multicolumn{4}{c}{$4<\xi<5$}\\
\hline
\multicolumn{2}{c|}{Jet \pt range (\GeVcns)} & 100--120 & 120--150 & 150--300 & 100--300\\
\hline

Jet energy resolution & 10--20\% smearing & 0.2--0.6\% & 0.7\% & 1.0--5.1\% & 1.7--2.5\%\\
Jet energy scale & 5\% shift & 0.21--2.9\% & 0.19--2.7\% & 0.18--2.5\% & 0.2--2.8\%\\
Tracking efficiency & Non-closure & \multicolumn{4}{c}{0.2--1.6\%}\\
Tracking efficiency & Centrality variation & \multicolumn{4}{c}{10\%}\\
Background bias &50\% of correction & \multicolumn{4}{c}{8.1--8.8\%}\\
Background procedure & Diff. between methods & \multicolumn{4}{c}{3.8--4.9\%}\\
Gluon jet cross-check & Non-closure & \multicolumn{4}{c}{1.1--4.2\%}\\
Quark jet cross-check & Non-closure & \multicolumn{4}{c}{1.5--4.5\%}\\
\hline
Total & & 19--26\% & 16--17\% & 17--23\% & 14--16\%\\
\end{scotch}
}
\end{table*}

Systematic uncertainties are primarily due to the tracking efficiency estimation, jet reconstruction, subtraction of the underlying background, and differences between MC simulations and data.

One systematic uncertainty arising from the tracking efficiency and rate of reconstructing misidentified (fake) tracks is
estimated using the Monte Carlo simulations.
This uncertainty is based on the comparison of the fragmentation function obtained by corrected, reconstructed
tracks with those obtained using the generator-level particles.
In both cases, the same (\ie matched) reconstructed and generated jets are used.
An additional 10\% tracking efficiency uncertainty is assigned to account for the effects found in a study
of the charge composition of reconstructed jets.
Specifically, the observed charged fraction of a jet (as determined using the PF objects within the jet) decreases
with in the more central collisions with a 10\% difference between central and peripheral
heavy-ion data (the latter being very similar to \pp data and the MC results).
The difference in charge fraction between central and peripheral MC events is less than 2\%.
It is possible that this change is due to physics. However, to be conservative, a value of 10\% is used as an estimate
of a possible tracking inefficiency in high multiplicity \PbPb events that is possibly not reproduced by \PYHYD.
The estimated uncertainty ranges from 0 for peripheral events to 10\% for the most central events.

The fragmentation function is also sensitive to uncertainties
in the jet reconstruction in two aspects: the smearing of jet energy due to resolution and the overall energy scale.
In order to estimate the effects of the jet energy resolution,
a fragmentation function is constructed from the same \PbPb data but with
the jet energy artificially modified by smearing with the additional jet energy resolution due to the underlying event.
Comparison of the fragmentation function with and without modified jet energies is used to estimate the impact of these underlying event fluctuations. The full difference found between the two fragmentation functions is assigned as a systematic uncertainty.
The uncertainty due to the jet energy scale is estimated by varying the energy of jets in the \PbPb data by 5\%.
This factor is determined using a quadratic sum of the following three contributions.
The uncertainty in the jet energy scale in \pp data was found to be 2--3\%~\cite{Chatrchyan:2011ds}.
Comparing jet energies in reconstructed \PYHYD events to the generator-level values showed a variation of about 1\%.
As will be shown in Section~\ref{section_Results}, the observed modification in the \PbPb fragmentation function is largely characterized by an excess of particles with $\pt < 3\GeVc$ ($\langle\pt\rangle = 1.4\GeVc$). The jet energy corrections used in the analysis are found using MC jets lacking these additional low-\pt particles.  The impact of this effect was studied by inserting extra particles in the $1 < \pt < 2\GeVc$ range to the \PYHYD simulated jets. The ratio of reconstructed over generator-level fragmentation functions is found to vary by about 4\% from the ratio found without extra embedded particles.
The impact of the uncertainties due to the jet energy scale and resolution are different since the scale uncertainty shifts all jet \pt values in the same direction and the \pt shift due to resolution can be larger than that due to the scale uncertainty. The resulting changes in the fragmentation function depend on both the bin width in $\pt^\text{jet}$ and the range of $\pt^\text{track}$ that corresponds to a given bin in $\xi$. Thus, the fragmentation function uncertainties are a somewhat complex function of $\pt^\text{jet}$ and $\xi$.

To estimate the uncertainty in the  underlying-event background subtraction, the fragmentation function analysis is repeated with an alternative background subtraction using mixed events.
This method estimates the background contribution to the fragmentation function using minimum bias \PbPb data. For each jet in a signal event, tracks around the ($\eta,\phi$) position of the jet are taken from a separate minimum bias event. The total \pt of these tracks constitutes the ``mixed-event'' background.
The full difference between the two background subtraction methods is quoted conservatively as the uncertainty in the fragmentation function due to background subtraction.
 Furthermore, due to the requirement of a minimum reconstructed jet energy, jets sitting on top of an upward background fluctuation are more
likely to be selected. In MC events, it is known precisely which particles are from the underlying event and which are from the hard scattering
signal, and therefore a correction for the impact of this effect on the default ($\eta$-reflected cone) background subtraction can be extracted.
It is possible that the  point-to-point underlying event fluctuations are slightly different between data and the MC.
To account for this possibility, half of the size of the applied correction is quoted as a systematic uncertainty.

Finally to estimate the effect of the signal jets in data having a different fragmentation pattern than jets in simulations,
different types of signal jets are embedded into simulated heavy-ion events. Systematic uncertainties are quoted based on the
difference between the reconstructed fragmentation functions with embedded gluon and quark jets.

The individual systematic uncertainties are added in quadrature to form the total systematic uncertainty. This is summarized in Table~\ref{table:JetFFSys}. The quoted systematic uncertainty band is cross-checked by repeating the complete analysis with \PYHYD corrections applied to reconstructed events in which quenched jets generated using {\textsc{pyquen}}~\cite{Lokhtin:2005px} are embedded. The reconstructed fragmentation function from this study is consistent with the generator truth within the quoted systematic uncertainties.
\section{Results}
\label{section_Results}
\begin{figure*}[tbp]
   \centering
      \includegraphics[width=\cmsFigWidth]{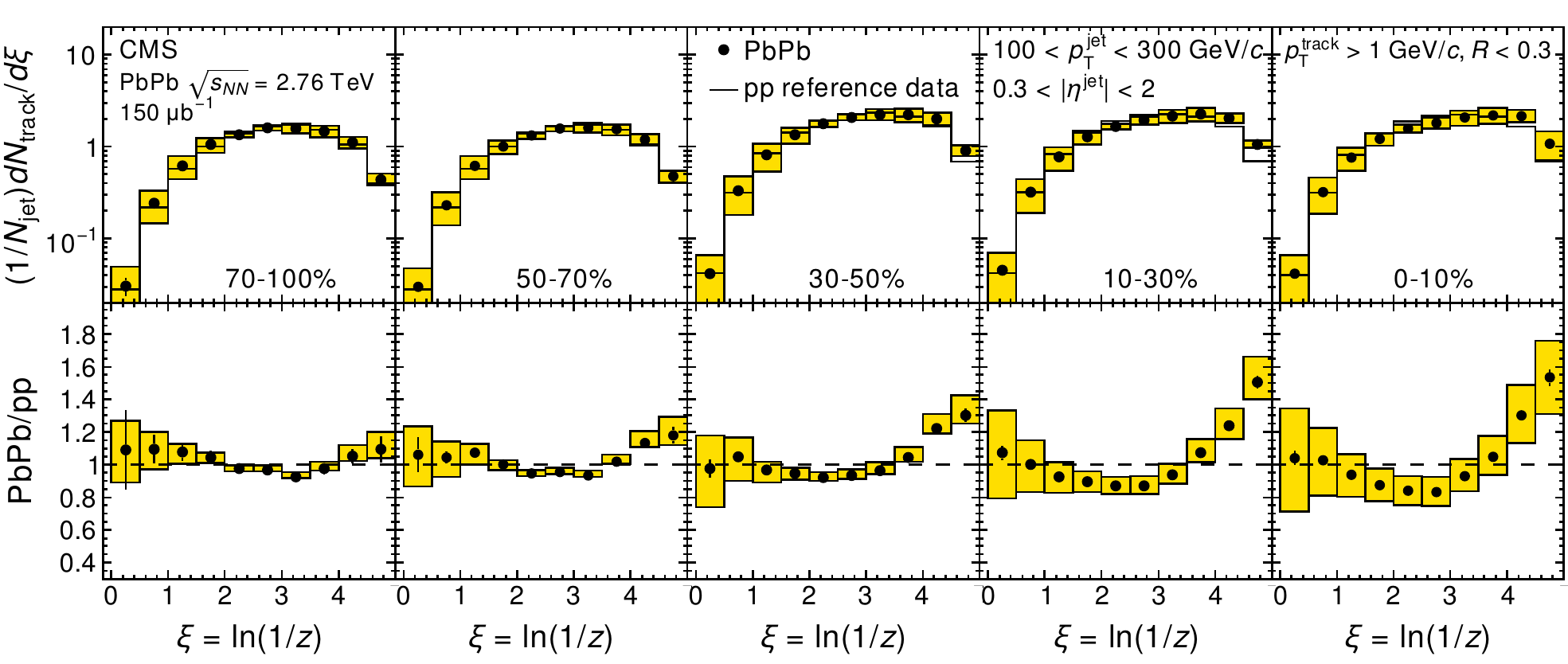}
      \caption{(Color online) (Top) The \PbPb fragmentation function in bins of centrality (increasing from left to right) overlaid with \pp reference data.
      Jets have $100 < \pt < 300\GeVc$, and tracks have $\pt > 1\GeVc$.
      (Bottom) The ratio of each \PbPb fragmentation function to its \pp reference. Error bars are statistical,
      and boxes show the systematic uncertainty.}
      \label{fig:FF_EtaRef_vs_Cent}
\end{figure*}

\begin{figure*}[tbp]
   \centering
      \includegraphics[width=\cmsFigWidth]{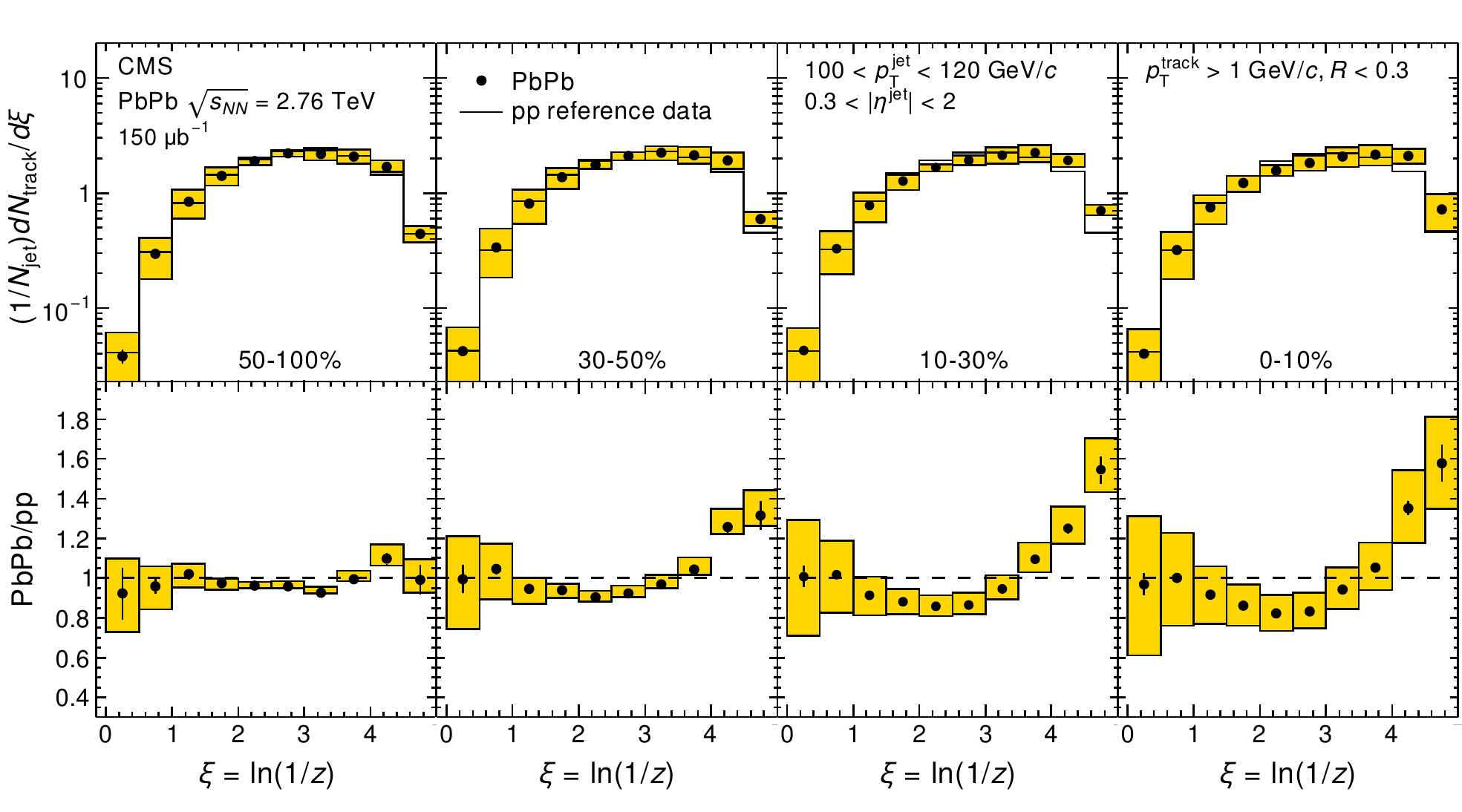}
      \caption{(Color online) (Top) The \PbPb fragmentation function in bins of centrality (increasing from left to right) overlaid with \pp reference data.
      Jets have $100 < \pt < 120\GeVc$, and tracks have $\pt > 1\GeVc$.
      (Bottom) The ratio of each \PbPb fragmentation function to its \pp reference. Error bars are statistical,
      and boxes show the systematic uncertainty.}
      \label{fig:FF_EtaRef_vs_Cent_100_120}
\end{figure*}

\begin{figure*}[tbp]
   \centering
      \includegraphics[width=\cmsFigWidth]{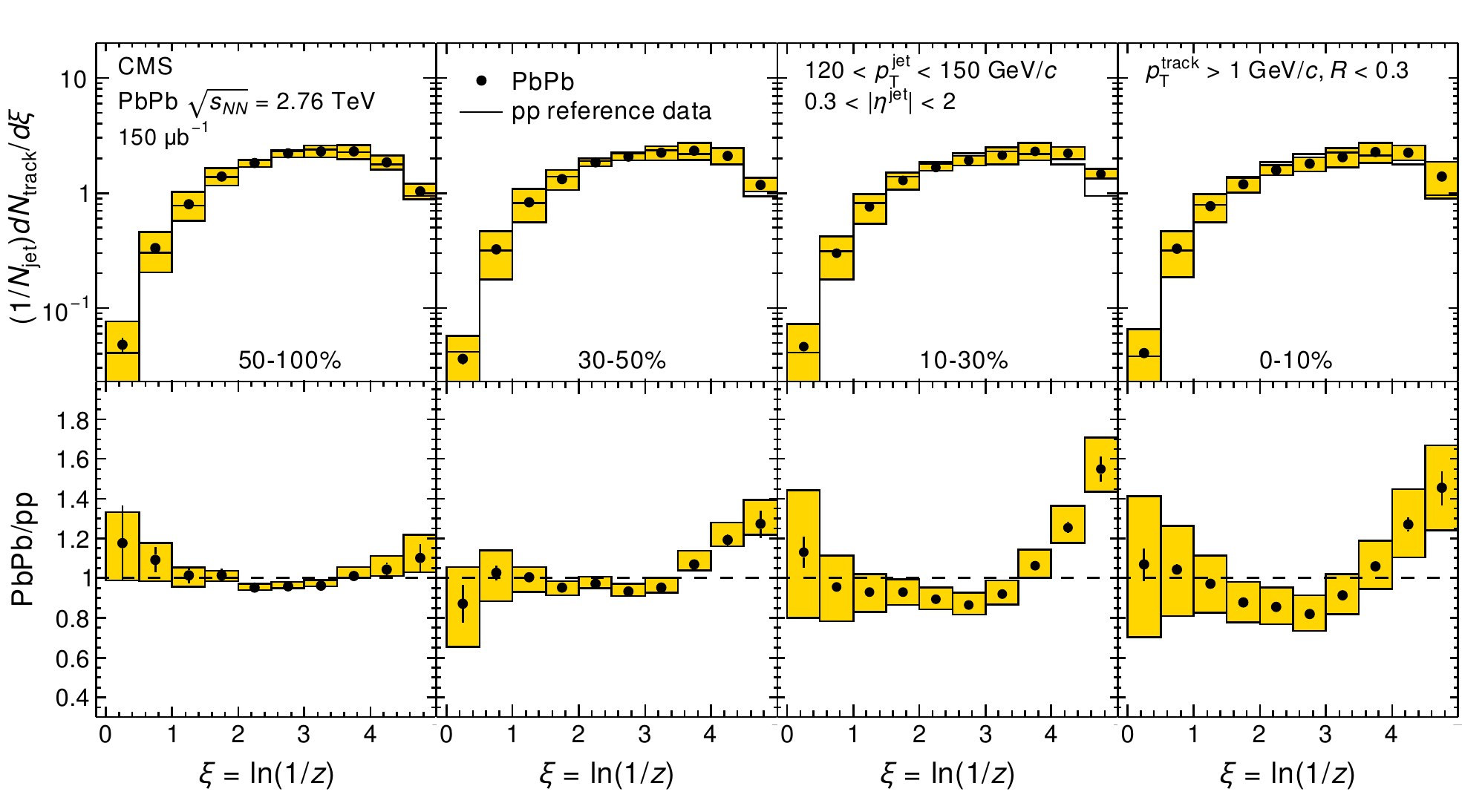}
      \caption{(Color online) (Top) The \PbPb fragmentation function in bins of centrality (increasing from left to right) overlaid with \pp reference data.
      Jets have $120 < \pt < 150\GeVc$, and tracks have $\pt > 1\GeVc$.
      (Bottom) The ratio of each \PbPb fragmentation function to its \pp reference. Error bars are statistical,
      and boxes show the systematic uncertainty.}
      \label{fig:FF_EtaRef_vs_Cent_120_150}
\end{figure*}

\begin{figure*}[tbp]
   \centering
      \includegraphics[width=\cmsFigWidth]{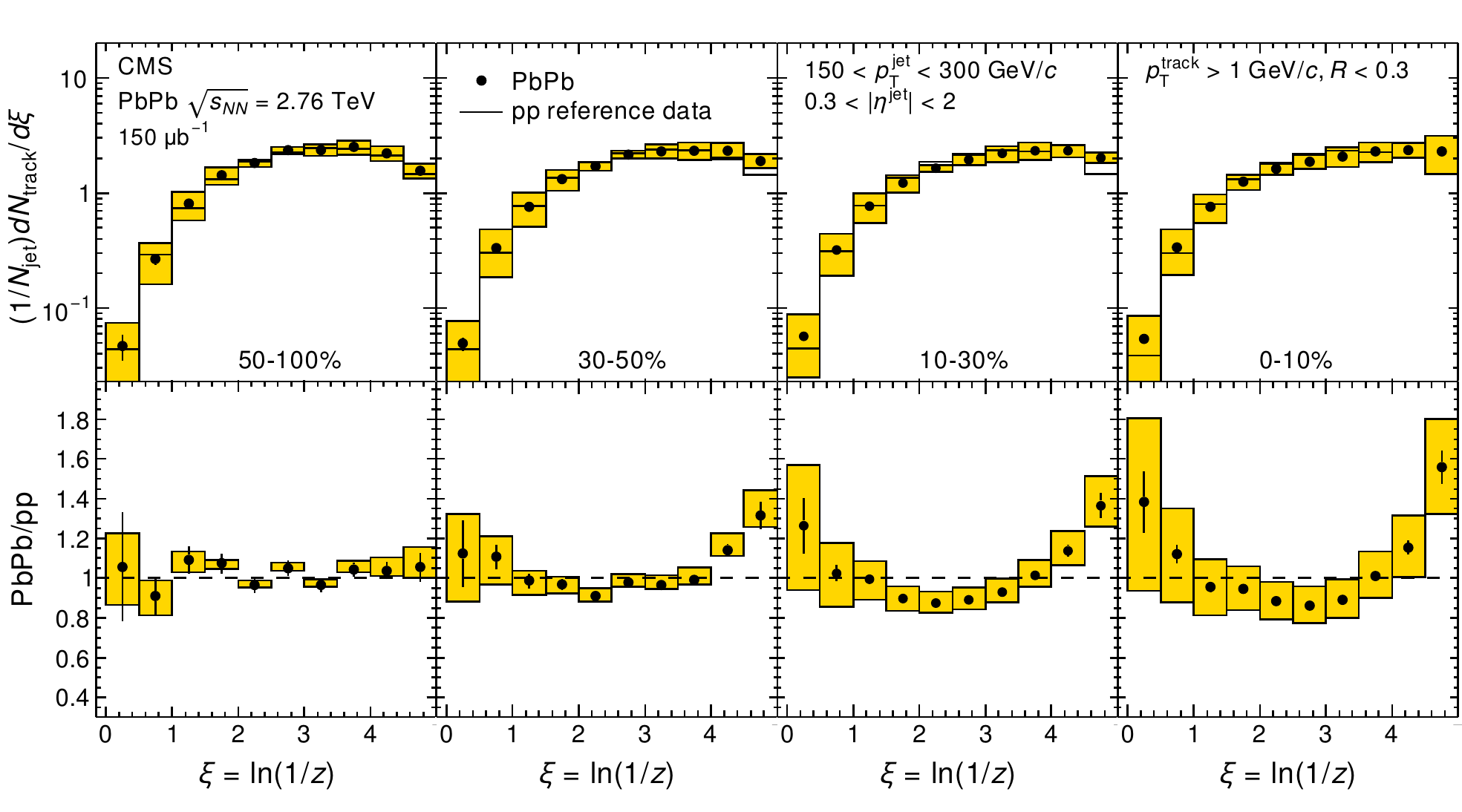}
      \caption{(Color online) (Top) The \PbPb fragmentation function in bins of centrality (increasing from left to right) overlaid with \pp reference data.
      Jets have $150 < \pt < 300\GeVc$, and tracks have $\pt > 1\GeVc$.
      (Bottom) The ratio of each \PbPb fragmentation function to its \pp reference. Error bars are statistical,
      and boxes show the systematic uncertainty.}
      \label{fig:FF_EtaRef_vs_Cent_150_300}
\end{figure*}

Figure~\ref{fig:FF_EtaRef_vs_Cent}
shows the fragmentation function reconstructed in \PbPb data and \pp reference for $100 < \pt^\text{jet} < 300\GeVc$ and tracks with \pt above 1\GeVc within a radius of 0.3 relative to the corresponding jet axis. For the \PbPb fragmentation function, the contribution from the underlying event is subtracted using the $\eta$ reflection method. For the \pp reference data, the corresponding jet distribution is first smeared with the additional \PbPb jet resolution due to the underlying event, and then re-weighted to match the jet \pt distribution in \PbPb data.  Figures~\ref{fig:FF_EtaRef_vs_Cent_100_120}--\ref{fig:FF_EtaRef_vs_Cent_150_300} show the same fragmentation function differentially in jet \pt, for $100 < \pt^\text{jet} < 120\GeVc$, $120 < \pt^\text{jet} < 150\GeVc$, and $150 < \pt^\text{jet} < 300\GeVc$, respectively.  In the higher statistics Figure \ref{fig:FF_EtaRef_vs_Cent}, we retained the same 50--70\% and 70--100\% centrality binning from \cite{Chatrchyan:2013kwa}.

It is clear from the results in Figs.~\ref{fig:FF_EtaRef_vs_Cent}--\ref{fig:FF_EtaRef_vs_Cent_150_300} that the modification of the fragmentation function of jets in \PbPb compared to those in \pp grows with increasingly central collision.
In the 50--100\% bin, the ratio of PbPb/pp is almost flat at unity within the systematic uncertainties, which means no modification.
However, a significant excess at high $\xi$ (low track \pt) is observed for more central events, combined with a depletion in the intermediate $\xi$.
In the most central 0--10\% collisions and for the lowest charged-particle
momenta studied, the PbPb/pp fragmentation function ratio rises to
$\sim$1.5.
This implies that for central collisions
the spectrum of particles in a jet has an enhanced contribution of soft particles compared to that from \pp collisions.  No significant dependence of the modification on $\pt^\text{jet}$ is observed within our current statistical and systematic uncertainties.

Possible sources of the observed modifications to the fragmentation function in central \PbPb collisions include a change in the fraction
of jets arising from either quarks or gluons, a change in the parton shower due to the medium \cite{Wang:2001ifa,Renk:2008pp,Renk:2009nz,Armesto:2009fj,Wang:2009qb}, or the presence of particles resulting
from the medium response. The fragmentation patterns of pure quarks and partons are predicted by \PYTHIA to be significantly different.
If traversing the medium has a bigger impact on gluons, as might be expected given their larger color charge compared to quarks, this
change in the admixture of the two parton types would change the measured fragmentation function. The distinction between the second
and third listed possibilities may or may not be valid depending on the specifics of the physical processes responsible for jet quenching.
More detailed experimental and theoretical analysis will be required to attempt to separate the influence of these, and possibly other,
contributions to the observed effects.

\begin{figure*}[tbp]
   \centering
      \includegraphics[width=\cmsFigWidth]{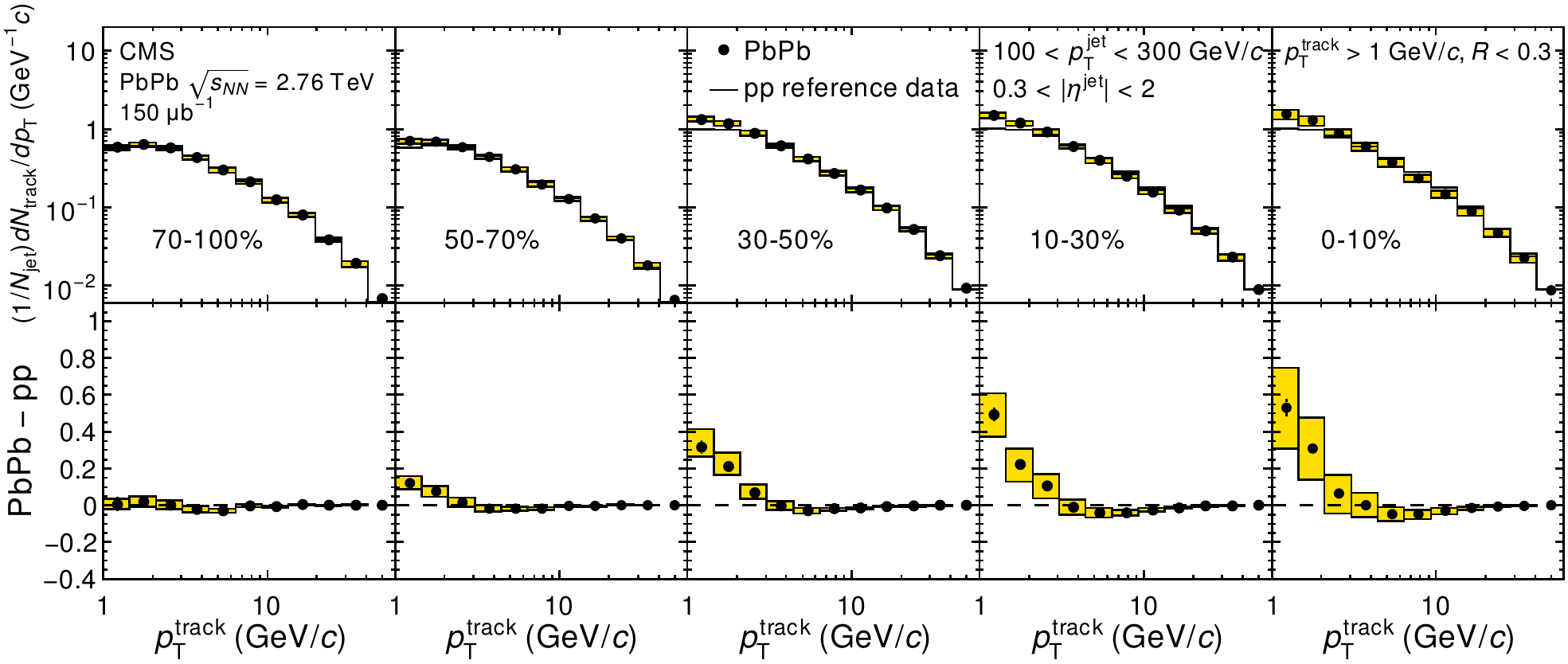}
      \caption{(Color online) The spectrum of tracks inside the cone of jets with $100 < \pt^\text{jet} < 300\GeVc$, as a function of track \pt, for \PbPb (with increasing centrality from left to right) and pp. Both the \PbPb and \pp results are background subtracted, in the same manner as for the fragmentation function. The bottom panels show the difference of \PbPb and \pp spectra, demonstrating an excess of low-\pt\ tracks in the \PbPb events. }
      \label{fig:jetFF_trackpt}
\end{figure*}

\begin{figure*}[tbp]
   \centering
      \includegraphics[width=\cmsFigWidth]{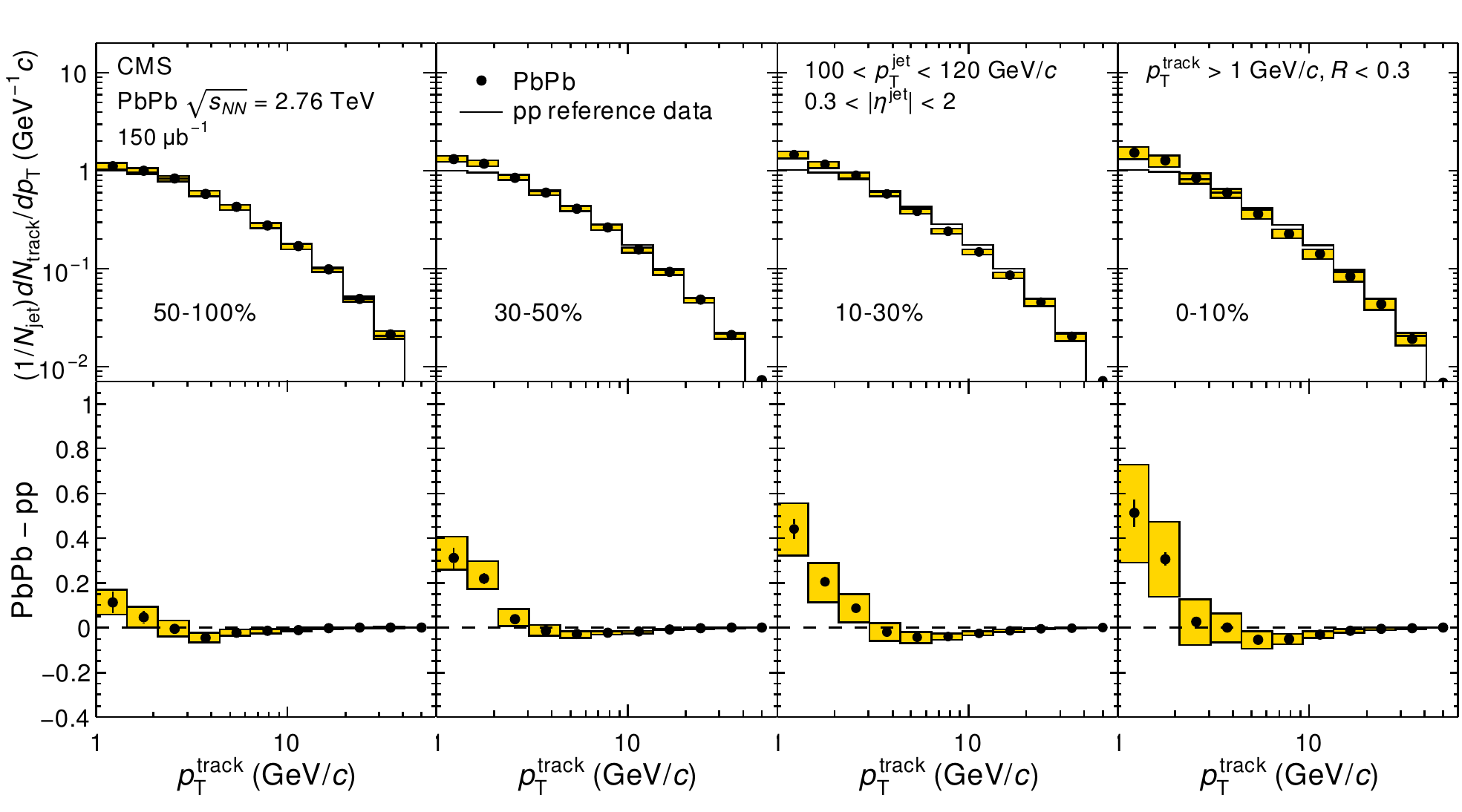}
      \caption{(Color online) The spectrum of tracks inside the cone of jets with $100 < \pt^\text{jet} < 120\GeVc$, as a function of track \pt, for \PbPb (with increasing centrality from left to right) and pp. Both the \PbPb and \pp results are background subtracted, in the same manner as for the fragmentation function. The bottom panels show the difference of \PbPb and \pp spectra, demonstrating an excess of low-\pt\ tracks in the \PbPb events.}
      \label{fig:jetFF_trackpt_100_120}
\end{figure*}

\begin{figure*}[tbp]
   \centering
      \includegraphics[width=\cmsFigWidth]{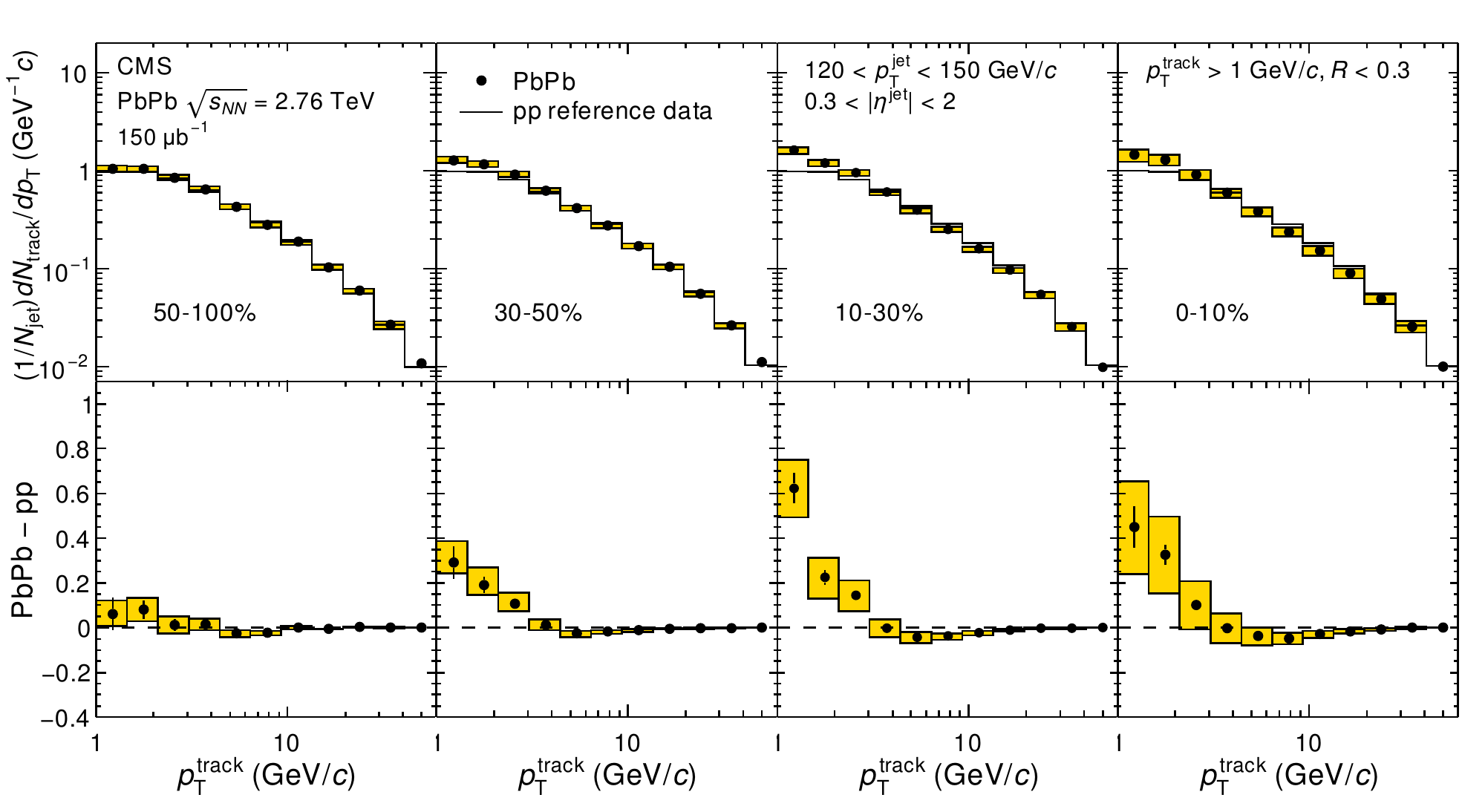}
      \caption{(Color online) The spectrum of tracks inside the cone of jets with $120 < \pt^\text{jet} < 150\GeVc$, as a function of track \pt, for \PbPb (with increasing centrality from left to right) and pp. Both the \PbPb and \pp results are background subtracted, in the same manner as for the fragmentation function. The bottom panels show the difference of \PbPb and \pp spectra, demonstrating an excess of low-\pt\ tracks in the \PbPb events. }
      \label{fig:jetFF_trackpt_120_150}
\end{figure*}

\begin{figure*}[tbp]
   \centering
      \includegraphics[width=\cmsFigWidth]{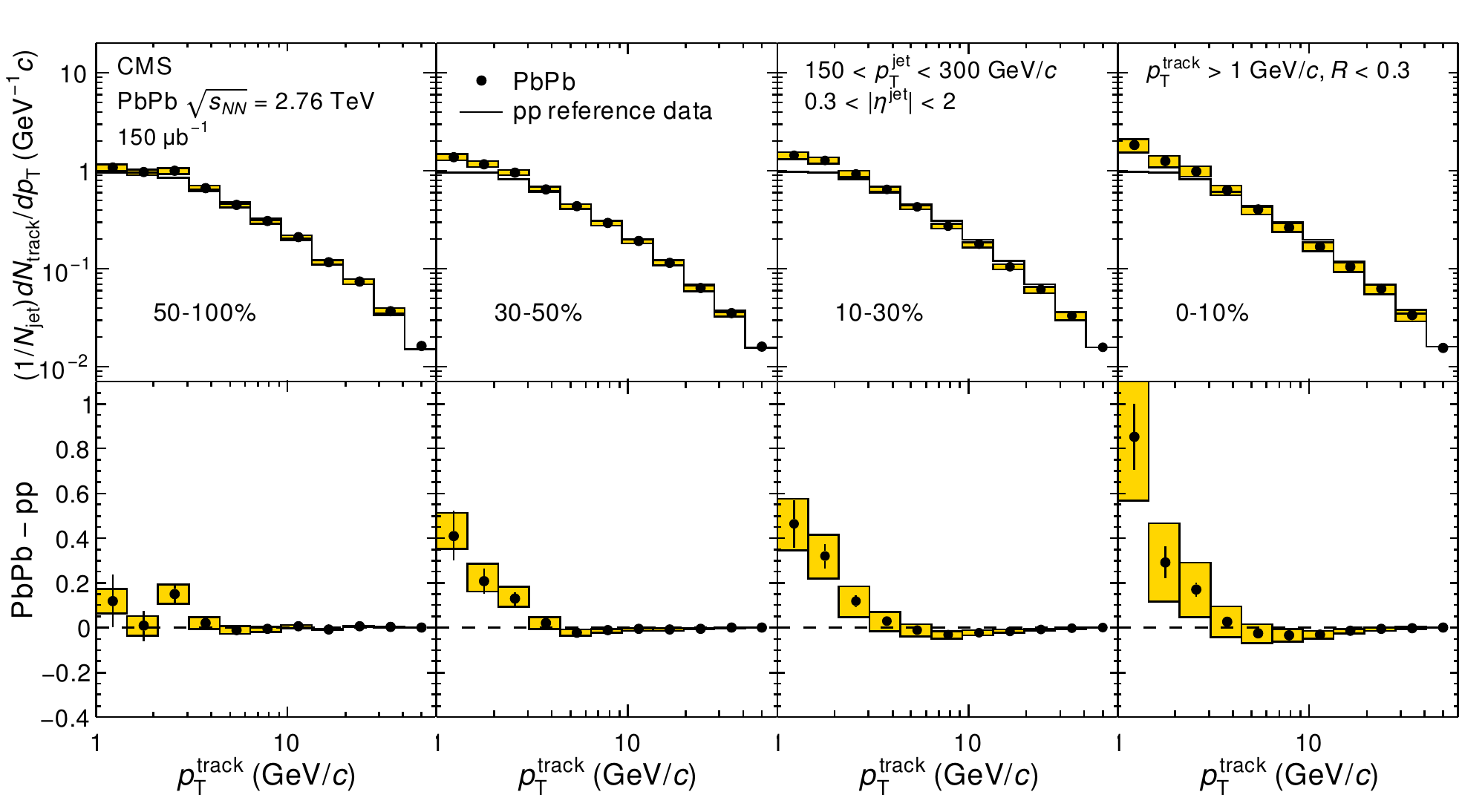}
      \caption{(Color online) The spectrum of tracks inside the cone of jets with $150 < \pt^\text{jet} < 300\GeVc$, as a function of track \pt, for \PbPb (with increasing centrality from left to right) and pp. Both the \PbPb and \pp results are background subtracted, in the same manner as for the fragmentation function. The bottom panels show the difference of \PbPb and \pp spectra, demonstrating an excess of low-\pt\ tracks in the \PbPb events. }
      \label{fig:jetFF_trackpt_150_300}
\end{figure*}

One can further investigate in which track \pt\ ranges the fragmentation function exhibits an excess by examining the \pt\ spectra for tracks inside the jet cone. These distributions are obtained with the same background subtraction described above.
Figure~\ref{fig:jetFF_trackpt} shows the spectra of tracks in the jet cone compared to \pp reference data.  In order to quantify the excess of tracks at a given \pt, the bottom panels show the difference of the two distributions, \pp subtracted from PbPb. Figures~\ref{fig:jetFF_trackpt_100_120}--\ref{fig:jetFF_trackpt_150_300} show the same differentially in $\pt^\text{jet}$, for $100 < \pt^\text{jet} < 120\GeVc$, $120 < \pt^\text{jet} < 150\GeVc$, and $150 < \pt^\text{jet} < 300\GeVc$, respectively. The excess that is observed at the high-$\xi$ region of the fragmentation function is localized at low-\pt\ for tracks (\pt below $\approx$3\GeVc). No pronounced jet \pt dependence of this excess is observed within the current statistical and systematic uncertainties.

Figures~\ref{fig:FF_EtaRef_vs_Cent}--\ref{fig:FF_EtaRef_vs_Cent_150_300} also show some evidence for a very weak suppression in the intermediate region, $\xi\approx 1.5$--3.0.
This corresponds to the weak suppression seen in Figs.~\ref{fig:jetFF_trackpt}--\ref{fig:jetFF_trackpt_150_300} around $\pt \approx 6\GeVc$.
Since summing the \pt of all included tracks cannot exceed the total jet \pt, it is impossible to have significantly more tracks in one \pt range without having fewer in another range.
However, fewer tracks are required at high-\pt to satisfy this summed-momentum restriction.
While some depletion is observed, the uncertainties in the current results preclude a precise determination of the \pt range from which the excess low-\pt tracks originate.

\section{Summary}

The fragmentation function of inclusive jets in \PbPb collisions at $\sNN= 2.76$\TeV
has been measured.
Jets were reconstructed using the anti-\kt algorithm with a distance parameter of 0.3.

For the analysis, inclusive jets with $\pt^{\text{jet}} > 100$\GeVc and $0.3 < \abs{\eta_\text{jet}} < 2$ were reconstructed using particle-flow  objects, which combine information from charged-particle tracking and calorimetry.
The jet fragmentation function in a cone of 0.3 was obtained using charged particles with $\pt > 1\GeVc$ and $\abs{\eta} < 2.4$, and as a function of collision
centrality for five centrality selections, 70--100\%, 50--70\%, 30--50\%, 10--30\%, and 0--10\%. The
uncorrelated contribution from the underlying event to the charged-particle distribution
in the cone was subtracted using an ``$\eta$-reflected cone'' method.
The fragmentation function in \PbPb\ collisions
was compared to measurements with the same selection in \pp\ collisions at the same
center-of-mass  energy. For this comparison,  a jet momentum smearing and reweighting
procedure was applied to obtain a proper pp-based reference.

For the 70--100\% most peripheral collisions, the fragmentation
function in \PbPb collisions agrees with that for the \pp reference. Going to more central
collisions, a significant modification of the
fragmentation function in \PbPb compared to \pp in the intermediate and
high $\xi$-region develops.
For charged particles in the region  of 1--3\GeVc, corresponding to $\xi$ above about 3.5, a clear rise in the  ratio
of \PbPb to \pp is observed.
In the intermediate \pt\ range of fragmentation products, $\xi$ between 2 and 3, evidence is seen for a small depletion in the ratio.
In the most central 0--10\% collisions and for the lowest charged
particle momenta studied, the PbPb/pp fragmentation function ratio rises to
$\approx$1.6. For the current jet kinematic range, no significant variation of the modification with jet \pt is observed within the statistical and systematic uncertainties.

By including charged particles at lower \pt than the previous CMS measurement,
a clear centrality dependent modification of the inclusive jet
fragmentation function in \PbPb collisions is now revealed.
As fragmentation at larger radii from the jet axis is dominated by low \pt particles, this is consistent with the
enhancement seen in \cite{Chatrchyan:2013kwa}.
The interplay between the modifications in the high-\pt and low-\pt parts of the fragmentation function provides constraints on models of medium induced energy loss, and opens up new avenues to understand the transport properties of the QGP.

\clearpage

\section{Acknowledgements}

\hyphenation{Bundes-ministerium Forschungs-gemeinschaft Forschungs-zentren} We congratulate our colleagues in the CERN accelerator departments for the excellent performance of the LHC and thank the technical and administrative staffs at CERN and at other CMS institutes for their contributions to the success of the CMS effort. In addition, we gratefully acknowledge the computing centres and personnel of the Worldwide LHC Computing Grid for delivering so effectively the computing infrastructure essential to our analyses. Finally, we acknowledge the enduring support for the construction and operation of the LHC and the CMS detector provided by the following funding agencies: the Austrian Federal Ministry of Science, Research and Economy and the Austrian Science Fund; the Belgian Fonds de la Recherche Scientifique, and Fonds voor Wetenschappelijk Onderzoek; the Brazilian Funding Agencies (CNPq, CAPES, FAPERJ, and FAPESP); the Bulgarian Ministry of Education and Science; CERN; the Chinese Academy of Sciences, Ministry of Science and Technology, and National Natural Science Foundation of China; the Colombian Funding Agency (COLCIENCIAS); the Croatian Ministry of Science, Education and Sport, and the Croatian Science Foundation; the Research Promotion Foundation, Cyprus; the Ministry of Education and Research, Estonian Research Council via IUT23-4 and IUT23-6 and European Regional Development Fund, Estonia; the Academy of Finland, Finnish Ministry of Education and Culture, and Helsinki Institute of Physics; the Institut National de Physique Nucl\'eaire et de Physique des Particules~/~CNRS, and Commissariat \`a l'\'Energie Atomique et aux \'Energies Alternatives~/~CEA, France; the Bundesministerium f\"ur Bildung und Forschung, Deutsche Forschungsgemeinschaft, and Helmholtz-Gemeinschaft Deutscher Forschungszentren, Germany; the General Secretariat for Research and Technology, Greece; the National Scientific Research Foundation, and National Innovation Office, Hungary; the Department of Atomic Energy and the Department of Science and Technology, India; the Institute for Studies in Theoretical Physics and Mathematics, Iran; the Science Foundation, Ireland; the Istituto Nazionale di Fisica Nucleare, Italy; the Korean Ministry of Education, Science and Technology and the World Class University program of NRF, Republic of Korea; the Lithuanian Academy of Sciences; the Ministry of Education, and University of Malaya (Malaysia); the Mexican Funding Agencies (CINVESTAV, CONACYT, SEP, and UASLP-FAI); the Ministry of Business, Innovation and Employment, New Zealand; the Pakistan Atomic Energy Commission; the Ministry of Science and Higher Education and the National Science Centre, Poland; the Funda\c{c}\~ao para a Ci\^encia e a Tecnologia, Portugal; JINR, Dubna; the Ministry of Education and Science of the Russian Federation, the Federal Agency of Atomic Energy of the Russian Federation, Russian Academy of Sciences, and the Russian Foundation for Basic Research; the Ministry of Education, Science and Technological Development of Serbia; the Secretar\'{\i}a de Estado de Investigaci\'on, Desarrollo e Innovaci\'on and Programa Consolider-Ingenio 2010, Spain; the Swiss Funding Agencies (ETH Board, ETH Zurich, PSI, SNF, UniZH, Canton Zurich, and SER); the Ministry of Science and Technology, Taipei; the Thailand Center of Excellence in Physics, the Institute for the Promotion of Teaching Science and Technology of Thailand, Special Task Force for Activating Research and the National Science and Technology Development Agency of Thailand; the Scientific and Technical Research Council of Turkey, and Turkish Atomic Energy Authority; the National Academy of Sciences of Ukraine, and State Fund for Fundamental Researches, Ukraine; the Science and Technology Facilities Council, UK; the US Department of Energy, and the US National Science Foundation.

Individuals have received support from the Marie-Curie programme and the European Research Council and EPLANET (European Union); the Leventis Foundation; the A. P. Sloan Foundation; the Alexander von Humboldt Foundation; the Belgian Federal Science Policy Office; the Fonds pour la Formation \`a la Recherche dans l'Industrie et dans l'Agriculture (FRIA-Belgium); the Agentschap voor Innovatie door Wetenschap en Technologie (IWT-Belgium); the Ministry of Education, Youth and Sports (MEYS) of the Czech Republic; the Council of Science and Industrial Research, India; the HOMING PLUS programme of Foundation for Polish Science, cofinanced from European Union, Regional Development Fund; the Compagnia di San Paolo (Torino); and the Thalis and Aristeia programmes cofinanced by EU-ESF and the Greek NSRF.

\bibliography{auto_generated}   

\cleardoublepage \appendix\section{The CMS Collaboration \label{app:collab}}\begin{sloppypar}\hyphenpenalty=5000\widowpenalty=500\clubpenalty=5000\textbf{Yerevan Physics Institute,  Yerevan,  Armenia}\\*[0pt]
S.~Chatrchyan, V.~Khachatryan, A.M.~Sirunyan, A.~Tumasyan
\vskip\cmsinstskip
\textbf{Institut f\"{u}r Hochenergiephysik der OeAW,  Wien,  Austria}\\*[0pt]
W.~Adam, T.~Bergauer, M.~Dragicevic, J.~Er\"{o}, C.~Fabjan\cmsAuthorMark{1}, M.~Friedl, R.~Fr\"{u}hwirth\cmsAuthorMark{1}, V.M.~Ghete, C.~Hartl, N.~H\"{o}rmann, J.~Hrubec, M.~Jeitler\cmsAuthorMark{1}, W.~Kiesenhofer, V.~Kn\"{u}nz, M.~Krammer\cmsAuthorMark{1}, I.~Kr\"{a}tschmer, D.~Liko, I.~Mikulec, D.~Rabady\cmsAuthorMark{2}, B.~Rahbaran, H.~Rohringer, R.~Sch\"{o}fbeck, J.~Strauss, A.~Taurok, W.~Treberer-Treberspurg, W.~Waltenberger, C.-E.~Wulz\cmsAuthorMark{1}
% ...
\vskip\cmsinstskip
% ...
1:~~Also at Vienna University of Technology, Vienna, Austria\\
2:~~Also at CERN, European Organization for Nuclear Research, Geneva, Switzerland\\
% ...

\end{sloppypar}
\end{document}